\documentclass[aps,epsf,twocolumn,superscriptaddress]{revtex4-1}
\usepackage{CJK,amsmath,amssymb,graphicx,epsfig,epstopdf,color,verbatim,ulem,braket,tabularx,float,bm}
\usepackage[colorlinks,linkcolor=blue,citecolor=blue,urlcolor=blue]{hyperref}

\begin{document}

\title{Ground States and Dynamical Properties of $S>1/2$ Quantum Heisenberg Model on the 1/5-Depleted Square Lattice}

\author{Jun-Han Huang}
\affiliation{Center for Neutron Science and Technology, State Key Laboratory of Optoelectronic Materials and Technologies, Guangdong Provincial Key Laboratory of Magnetoelectric Physics and Devices, School of Physics, Sun Yat-sen University, Guangzhou, 510275, China}
\affiliation{BGI-Shenzhen, Shenzhen 518083, China}

\author{Zenan Liu}
\affiliation{Center for Neutron Science and Technology, State Key Laboratory of Optoelectronic Materials and Technologies, Guangdong Provincial Key Laboratory of Magnetoelectric Physics and Devices, School of Physics, Sun Yat-sen University, Guangzhou, 510275, China}
	
\author{Han-Qing Wu}
\email{wuhanq3@mail.sysu.edu.cn}
\affiliation{Center for Neutron Science and Technology, State Key Laboratory of Optoelectronic Materials and Technologies, Guangdong Provincial Key Laboratory of Magnetoelectric Physics and Devices, School of Physics, Sun Yat-sen University, Guangzhou, 510275, China}

\author{Dao-Xin Yao}
\email{yaodaox@mail.sysu.edu.cn}
\affiliation{Center for Neutron Science and Technology, State Key Laboratory of Optoelectronic Materials and Technologies, Guangdong Provincial Key Laboratory of Magnetoelectric Physics and Devices, School of Physics, Sun Yat-sen University, Guangzhou, 510275, China}

\begin{abstract}
We study the $S>1/2$ antiferromagnetic Heisenberg model on the 1/5-depleted square lattice as a function of the ratio of the intra-plaquette coupling to the inter-plaquette coupling. Using stochastic series expansion quantum Monte Carlo simulations, we numerically identify three quantum phases, including the dimer phase, N\'eel phase and plaquette valence bond solid phase. We also obtain the accurate quantum critical points that belong to the O(3) universality class using the large-scale finite-size scaling. Most importantly, we study the dynamic spin structure factors of different phases, which can be measured by inelastic neutron scattering experiments. The low-energy excitations can be explained as triplons in the dimer phase and plaquette valence bond solid phase. While in the N\'eel phase, the more prominent magnon mode can be found as the spin magnitude increases. Furthermore, we find a broader continuum at smaller $S$, which may be the dynamical signature of nearly deconfined spinon excitations.
\end{abstract}

\date{\today}
\maketitle

\section{INTRODUCTION}
\begin{figure}[hbt]
	\centering
	\includegraphics[width=0.45\textwidth]{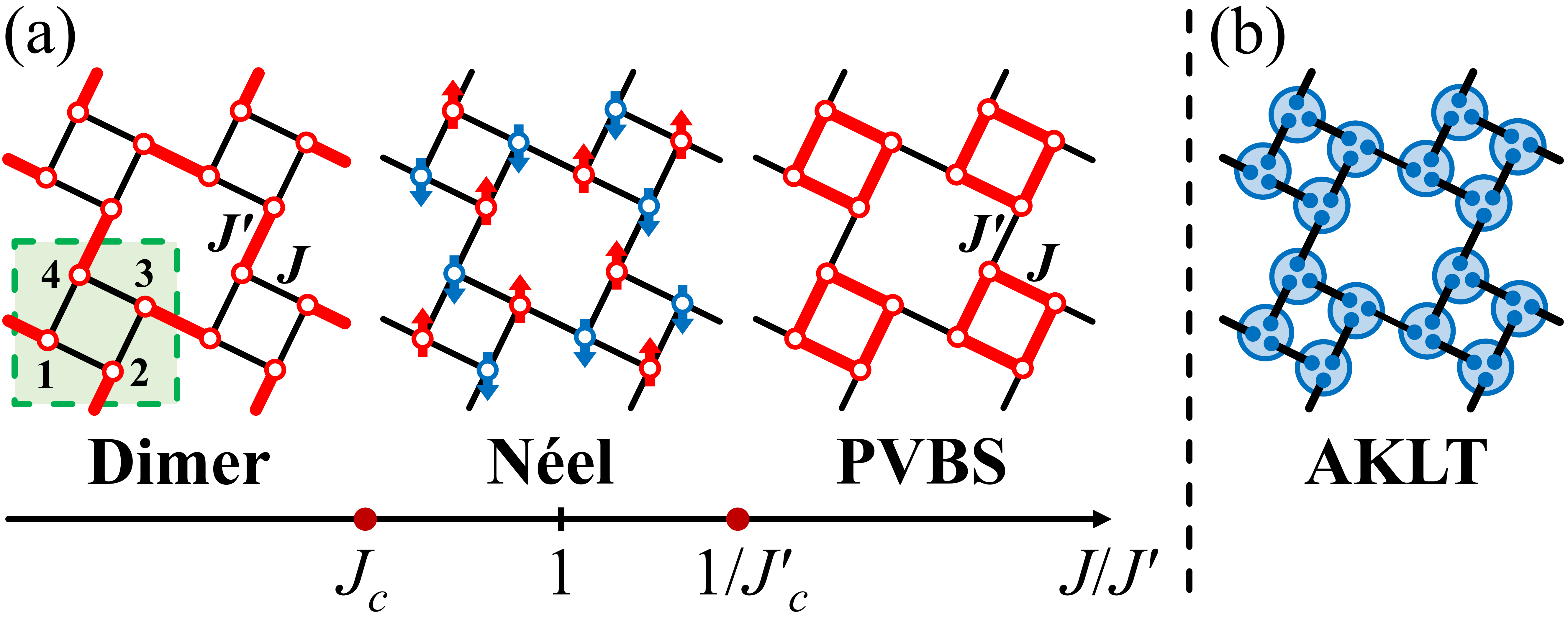}
	\includegraphics[width=0.45\textwidth]{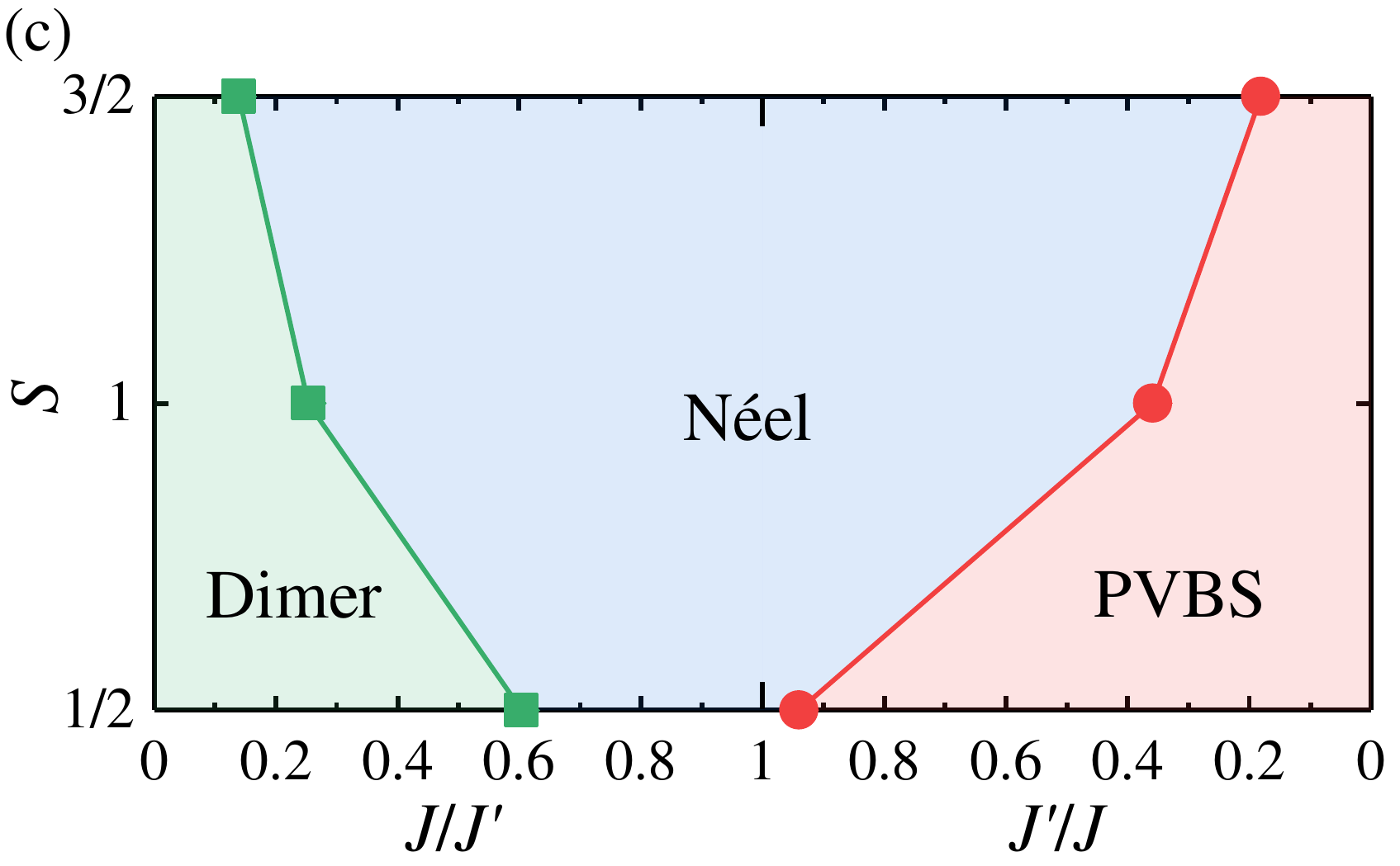}
	\caption{(a) The schematic phase diagram of the antiferromagnetic Heisenberg model on the 1/5-depleted square lattice, including two VBS phases and one N\'eel phase in between. $J$ and $J'$ are the intra-plaquette and inter-plaquette interactions, respectively. The green dashed square shows a unit cell with the sublattice labels $\alpha=1, 2, 3, 4$. (b) Illustration of the spin-3/2 AKLT state on the 1/5-depleted square lattice. Each spin-3/2 physical particle (blue large circles) can be regarded as the symmetric subspace of three virtual spin-1/2 particles (blue small circles). (c) The phase boundaries of the Heisenberg model with different spin magnitudes $S=1/2, 1, 3/2$. We set $J=1$ when $J>J'$, and set $J'=1$ when $J<J'$ in turn.
	}
	\label{phase}
\end{figure}

1/5-depleted square lattice has been found in some real materials, such as the compound $\mathrm{CaV_4O_9}$ \cite{Iwase1996, Starykh1996,Ueda1996,PhysRevLett.76.3822,PhysRevLett.77.3633} and the iron-based superconductor $\mathrm{ K_{0.8}Fe_{1.6}Se_{2}}$ \cite{PhysRevB.82.180520,PhysRevB.83.060512,Krzton_Maziopa_2011,Wang_2011,Fang_2011,PhysRevB.83.233205,PhysRevB.84.094451,CPhysLett.28.086104,Miaoyin2011, PhysRevB.84.140506, PhysRevLett.122.087201,RevModPhys.87.855,PhysRevLett.110.146402}. To understand the magnetic properties of these materials, we can study the Heisenberg model with competing intra-plaquette, inter-plaquette and other range exchange interactions. Among that, the competition of two unfrustrated intra-plaquette and inter-plaquette exchange interactions can induce quantum phase transitions between the N\'eel ordered phase and two valence-bond-solid phases [See Fig. \ref{phase}{\color{blue}(a)}], which can be handled by large-scale quantum Monte Carlo (QMC) simulation. Despite the well-studied of $S=1/2$ case, it is still worth accurately estimating the quantum critical points and studying its dynamical properties of this unfrustrated model in the higher-spin case. It is worth mentioning that the square-octagon lattice is topologically equivalent to the 1/5-depleted square lattice. The former lattice can be realized in the carbon-based material that can host superconductivity with doping \cite{PhysRevB.99.184506, ComMaterSci.110, PhysRevB.102.174509}.

Here, we briefly review some previous studies of the $S=1/2$ Heisenberg model on the 1/5-depleted square lattice \cite{PhysRevLett.76.3822,Schwandt2009, PhysRevB.98.140403, PhysRevB.105.014418, Yamada2014, Yanagi2014, Khatami2014, SciRep.4.6918, Wu2015}. When the ratio of the intra-plaquette coupling $J$ to the inter-plaquette coupling $J'$ tends to zero, the ground state of this model belongs to a dimer phase. And it can be effectively projected into a total $S=2$ Affleck-Kennedy-Lieb-Tasaki (AKLT) state on the square lattice \cite{PhysRevLett.118.087201}. Whereas the system forms a plaquette valence-bond-solid (PVBS) phase in the limit $J/J'\to\infty$. In between two gapped VBS phases, there is an intermediate N\'eel ordered phase, and two quantum critical points were estimated to be $J/J'=0.603520(10)$ and $J/J'=1.064382(13)$, respectively \cite{PhysRevLett.118.087201}. Thus, the slightly stronger intra-plaquette coupling or the frustrating next-nearest-neighbor interaction gives rise to the gapped PVBS phase, which has been used to explain the spin gap observed in the compound $\mathrm{CaV_4O_9}$ \cite{PhysRevLett.76.3822,Ueda1996,PhysRevLett.77.3633}.

For higher-spin case, the spin-$S$ AKLT state can be realized when the lattice coordination number $z$ and the spin quantum number $S$ satisfy the relation $z=2S/n$, where $n$ is a positive integer number \cite{AKLT1987, CommunMathPhys.115, PhysRevB.97.054412}. Importantly, the AKLT states on most of two-dimensional Archimedean lattices are universal resources for measurement based quantum computation \cite{Gross2007, Brennen2008, XChen2010,TCWei2011,PhysRevLett.124.177203,PhysRevA.92.012310,PhysRevA.70.060302,PhysRevA.88.062307,NatPhys.5.19}. The $S=3/2$ AKLT model on the honeycomb is a weak symmetry protected topological (SPT) phase that is protected by translational symmetry rather than the on-site symmetry \cite{PhysRevB.84.245128, PhysRevB.88.205124}, and it can be probed effectively by the strange correlator \cite{PhysRevLett.112.247202,PhysRevB.90.115157,PhysRevB.93.245141}. The $S=3/2$ model on 1/5-depleted square lattice also satisfies the relation $z=2S/n$ and can form AKLT phase shown in Fig. \ref{phase}{\color{blue}(b)}, in which each spin-3/2 physical particle can divide into three virtual spin-1/2 degrees of freedom and each two neighboring virtual spin-1/2 particles belonging to different physical particles can form a singlet. However, previous study has suggested that the AKLT model and the Heisenberg model may be not in the same phase on a trivalent lattice, unlike the one-dimensional chain \cite{CommunMathPhys.115,NewJPhys.14.013023,PhysRevB.94.165130,PhysRevX.10.021034,PhysRevB.92.024413}. The $S=3/2$ AKLT model on the 1/5-depleted square lattice that contains the biquadratic and bicubic terms is left for future study. In this work, we mainly study the $S>1/2$ antiferromagnetic Heisenberg model on the 1/5-depleted square lattice with only nearest-neighbor interactions by using large-scale QMC simulations.
We study the evolutions of phase boundaries and dynamical properties as the spin magnitude $S$ increases. Our numerical results can help to understand the magnetic properties of Mott insulator with multiorbitals on the 1/5-depleted square lattice or its topologically equivalent square-octagon lattice.

The rest of the paper is organized as follows. In Sec. \ref{model}, we introduce the Hamiltonian and the QMC methods for the higher-spin case. In Sec. \ref{result}, we study the phase boundaries based on the finite-size scaling hypothesis at criticality (see Fig. \ref{phase}), and we also study the dynamic spin structure factor of different phases by stochastic analytic continuation of QMC data. And the Sec. \ref{conclu} gives our final conclusion.

\section{MODEL AND METHODS}
\label{model}
We study the $S>1/2$ antiferromagnetic Heisenberg model on a 1/5-depleted square lattice, which is topologically equivalent to a square-octagon lattice. The Hamiltonian is expressed as,
\begin{equation}
\begin{aligned}
H=J\sum_{\langle ij \rangle}\bm{S}_i\cdot\bm{S}_j+J'\sum_{\langle ij \rangle'}\bm{S}_i\cdot\bm{S}_j,
\end{aligned}
\label{hamiltonian}
\end{equation}
where $\bm{S}_i$ denotes the spin-$S$ operator on each site $i$, $\langle ij \rangle$ denotes nearest-neighbor sites on the intra-plaquette bonds and $\langle ij \rangle'$ denotes the inter-plaquette bonds. $J, J'>0$ are the intra-plaquette and inter-plaquette antiferromagnetic couplings, respectively. For simplicity, we set $J'=1$ for $J\le J'$, and $J=1$ for $J>J'$ in the whole paper. Thus, the two potential quantum critical points can be expressed as $J_c$ ($J'=1$) and $J'_c$ ($J=1$), respectively.

To obtain the phase boundaries of the Heisenberg model on the 1/5-depleted square lattice, we employ sign-free QMC simulations based on the stochastic series expansion, which has been described in detail in Refs. \cite{Sandvikbook,PhysRevB.59.R14157,PhysRevE.66.046701,PhysRevB.66.134407}. Here, we briefly summarize two important update schemes used in the higher-spin case, comparing with the standard spin-1/2 one.

The first one is diagonal updates. Because of the non-uniform coupling strengths on the inter-plaquette bonds and intra-plaquette bonds, we consider the following acceptance probabilities to satisfy detailed balance:
\begin{equation}
\begin{aligned}
P([0,0]_p\rightarrow [1,b]_p)&={\rm min}\left[\frac{J_bN_b\beta \langle\alpha(p)|H_{1,b}|\alpha(p) \rangle}{L_n-n}, 1\right], \\
P([1,b]_p\rightarrow [0,0]_p)&={\rm min}\left[\frac{L_n-n+1}{J_bN_b\beta \langle\alpha(p)|H_{1,b}|\alpha(p) \rangle}, 1\right].
\end{aligned}
\label{accept}
\end{equation}
Here $L_n$ is the cut-off of the operator string, and $n$ is the number of the non-unit operators. And $J_b$ represents $J$ and $J'$ on the intra-plaquette and inter-plaquette bonds, respectively. More detailed explanations can be found in Ref. \cite{Sandvikbook}.

The second one is operator-loop updates. For the higher-spin case, even without external magnetic field and anisotropy, there are four possible paths through the vertices, including bounce, continue-straight, switch-and-reverse and switch-and-continue processes \cite{PhysRevE.66.046701}. Firstly, the operator-loop starts at a random position on one of the vertices, and the spin state on this position is changed to one of other $2S$ possible states. Next, according to the detailed balance condition, the paths are chosen with a probability proportional to the vertex weight. This procedure is repeated until the loop reaches the initial position meanwhile with a same spin state.

\section{NUMERICAL RESULTS}
\label{result}
We mainly explore the $S>1/2$ antiferromagnetic Heisenberg model on a 1/5-depleted square lattice with periodic boundary condition and the inverse temperature $\beta=1/T=\sqrt{N}$ (i.e., $\beta=2L$) unless specifically mentioned. Here, $L$ represents the linear number of the unit cells as shown in Fig. \ref{phase}{\color{blue}(a)}.

\subsection{The spin-3/2 case}
Firstly, we study the phase boundaries of the $S=3/2$ Heisenberg model on the 1/5-depleted square lattice. And two dimensionless quantities, the $L$-normalized uniform magnetic susceptibility $\chi L$ and the spin stiffness $\rho_s L$ are used to detect the quantum critical points \cite{PhysRevB.99.174434,PhysRevLett.121.117202}. The uniform magnetic susceptibility and the spin stiffness are expected to scale as $\chi \sim L^{z-d}$ and $\rho_s \sim L^{2-z-d}$ at the quantum critical points. Here, $d=2$ is the spatial dimension and $z=1$ is the dynamic critical exponent due to the O(3) universality class in this model. Thus, the two dimensionless quantities $\chi L$ and $\rho_s L$ are expected to be fixed values for different linear lattice sizes $L$ at the quantum critical points.

The uniform magnetic susceptibility $\chi$ is defined as
\begin{equation}
\begin{aligned}
\chi=\frac{\beta}{N} \left\langle\left(\sum_{i=1}^N S_i^z\right)^2\right\rangle,
\end{aligned}
\label{chi}
\end{equation}
and the $x$ direction of spin stiffness $\rho_s^x$ can be obtained from
\begin{equation}
\begin{aligned}
\rho_s^x=\frac{3}{2\beta N}\langle(N_x^+-N_x^-)^2\rangle,
\end{aligned}
\label{rhos}
\end{equation}
where $N_x^+$ and $N_x^-$ are the total number of off-diagonal operators transporting spin along the positive and negative $x$ direction, respectively. And the spin stiffness $\rho_s^y$ in the $y$ direction has the same definition as Eq. (\ref{rhos}). In practice, we extract spin stiffness from $\rho_s=(\rho_s^x+\rho_s^y)/2$ on the topologically equivalent square-octagon lattice with spatial isotropy.

\begin{figure}[tb]
	\centering
	\includegraphics[width=0.45\textwidth]{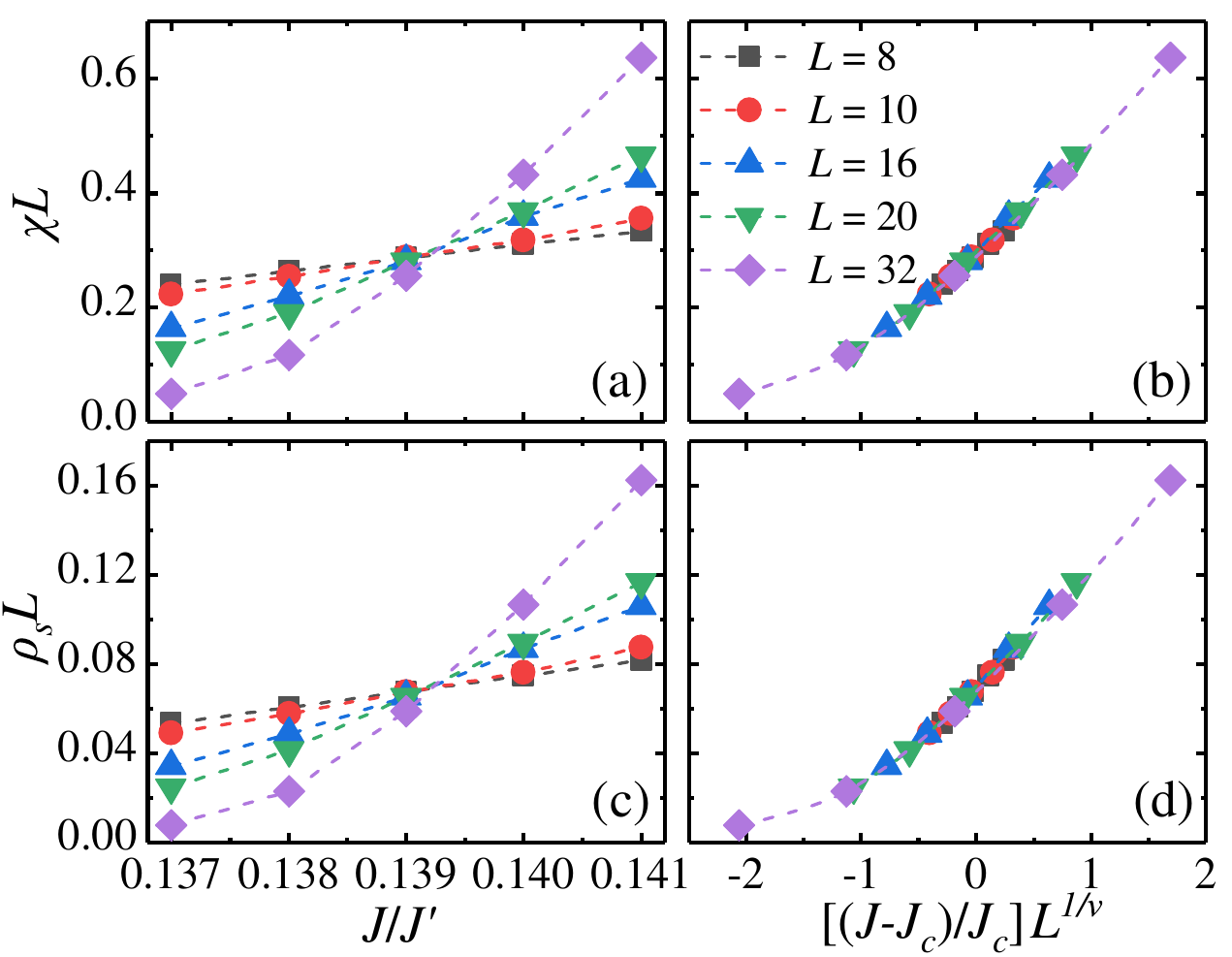}
	\caption{(a),(b) The uniform magnetic susceptibility multiplied by $L$ and (c),(d) the spin stiffness multiplied by $L$ of the $S=3/2$ antiferromagnetic Heisenberg model on the 1/5-depleted square lattice with $L=8, 10, 16, 20, 32$ near the quantum critical point $J_c$. The results are obtained from QMC simulations. Data collapses are achieved for a fixed standard O(3) value $1/\nu =1.406$ and $J_c=0.1392(2)$.}
	\label{s15qcp1}
\end{figure}

To obtain the quantum critical points, we calculate the dimensionless quantities $\chi L$ and $\rho_s L$ with different linear sizes $L=8, 10, 16, 20, 32$. As shown in Figs. \ref{s15qcp1}{\color{blue}(a)} and \ref{s15qcp1}{\color{blue}(c)}, the uniform magnetic susceptibility $\chi L$ and the spin stiffness $\rho_s L$ measured on different $L$ both roughly cross each other at the first critical point $J_c$, which confirms that a continuous quantum phase transition occurs between a dimer phase and a N\'eel phase \cite{PhysRevB.96.115160}. When the coupling ratio $J/J'<J_c$, the uniform magnetic susceptibility $\chi L$ gradually goes to zero as the linear size $L$ is increased. And the gapped dimer phase emerges as expected.

According to the finite-size scaling hypothesis at criticality, the dimensionless quantities satisfy the following form \cite{PhysRevB.79.054412,Sandvikbook}:
\begin{equation}
\begin{aligned}
Q(t,L)\sim f_Q(tL^{1/\nu}),
\end{aligned}
\label{fssa}
\end{equation}
where $t=(J-J_c)/J_c$ for $J\le J'$ is the reduced coupling and $\nu$ is the correlation length exponent. Here, we try to fix the correlation length exponent $\nu$ at the well-known standard O(3) value $1/\nu =1.406$ in the finite-size scaling analysis \cite{PhysRevB.65.144520}. Thus, we can proceed to perform data collapses in order to get a better estimate of the critical point $J_c$ based on Eq. (\ref{fssa}). As shown in Figs. \ref{s15qcp1}{\color{blue}(b)} and \ref{s15qcp1}{\color{blue}(d)}, we present the results of the dimensionless quantities $\chi L$ and $\rho_s L$ versus the reduced quantity $tL^{1/\nu}$. The good data collapses of $\chi L$ and $\rho_s L$ are achieved at $J_c=0.1392(2)$ when we set $J'=1$. Therefore, with the knowledge of three-dimensional O(3) universality class as expected, we get the accurate estimation of the first quantum critical point that is $J_c=0.1392(2)$.

\begin{figure}[tb]
	\centering
	\includegraphics[width=0.45\textwidth]{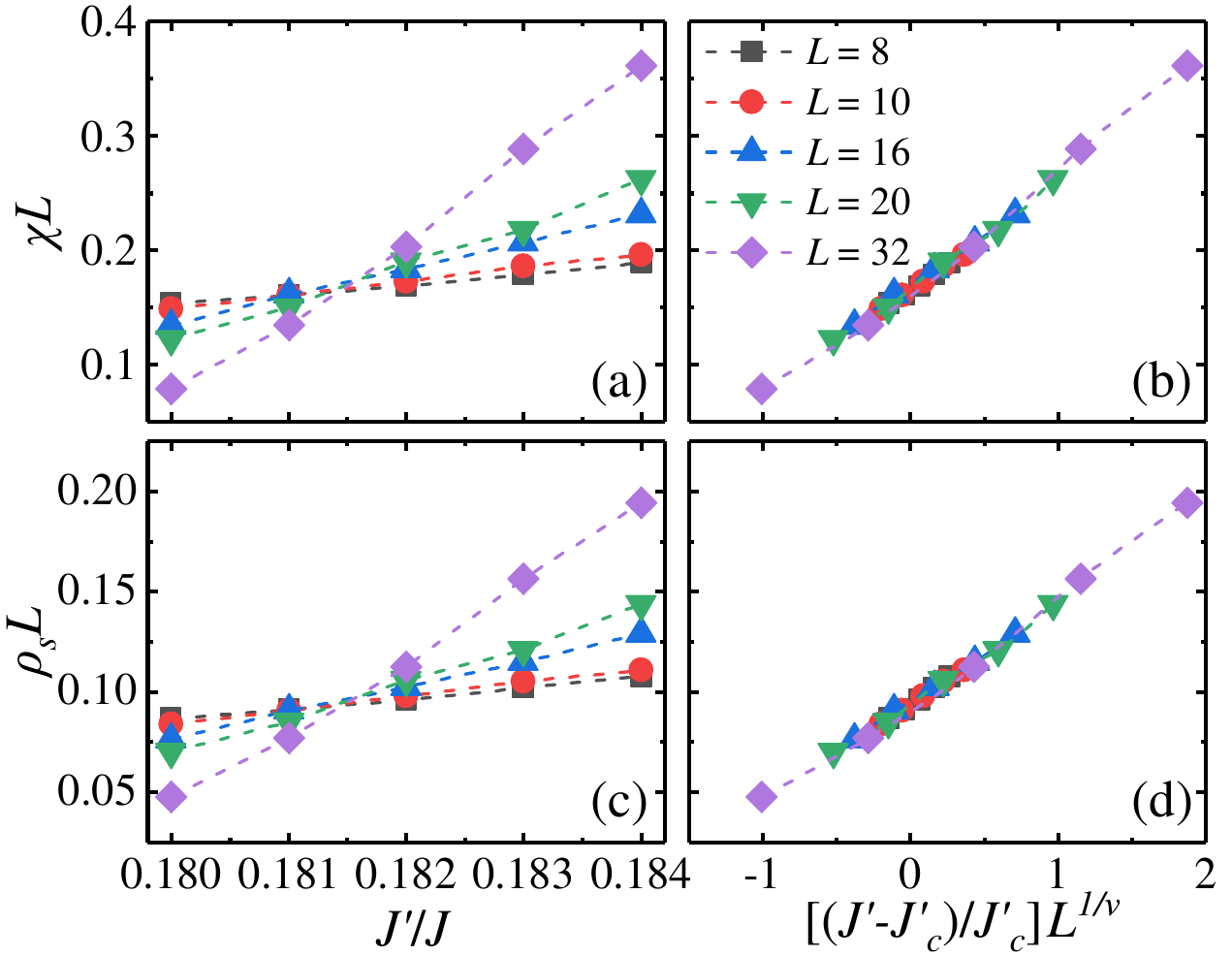}
	\caption{(a),(b) The uniform magnetic susceptibility multiplied by $L$ and (c),(d) the spin stiffness multiplied by $L$ of the $S=3/2$ antiferromagnetic Heisenberg model on the 1/5-depleted square lattice with $L=8, 10, 16, 20, 32$ near the quantum critical point $J'_c$. The results are obtained from QMC simulations. Data collapses are achieved for a fixed standard O(3) value $1/\nu =1.406$ and $J'_c=0.1814(4)$.}
	\label{s15qcp2}
\end{figure}

Next, we focus on the second quantum critical point in the phase diagram of the $S=3/2$ Heisenberg model on the 1/5-depleted square lattice. The uniform magnetic susceptibility $\chi L$ and the spin stiffness $\rho_s L$ with different $L$ cross each other again at the other critical point $J'_c$ (see Fig. \ref{s15qcp2}). Here we would like to emphasize that the second quantum critical point occurs at $J>J'$ instead of $J\le J'$. Moreover, a good data collapse of $\chi L$ and $\rho_s L$ with the accurate value $1/\nu=1.406$ for the O(3) universality class are obtained based on finite-size scaling as shown in Figs. \ref{s15qcp2}{\color{blue}(b)} and \ref{s15qcp2}{\color{blue}(d)}. The quantum critical point of this continuous phase transition is estimated to be $J'_c=0.1814(4)$ when we set $J=1$. And the disordered and gapped PVBS phase can be found when the ratio of the intra-plaquette coupling $J'$ to the inter-plaquette coupling $J$ is less than $0.1814(4)$.

Therefore, two quantum critical points of the $S=3/2$ Heisenberg model on the 1/5-depleted square lattice are obtained accurately to be $J_c=0.1392(2)$ and $J'_c=0.1814(4)$ using finite-size scaling. In Fig. \ref{phase}{\color{blue}(c)}, we show the phase diagram of this model versus the coupling ratios $J/J'$ for $J\le J'$ and $J'/J$ for $J>J'$, respectively. And the dimer phase and the PVBS phase represent the spin singlets formed on the inter-plaquette bonds and plaquettes, respectively \cite{PhysRevB.103.094421}. What's more, in the dimer phase, the spin-3/2 singlets are expected on the dimer bonds, and an effective gapless spin-3/2 chain is speculated to be formed by the dangling spins on the open edge similar to the spin-1/2 case \cite{PhysRevLett.118.087201,PhysRevLett.120.235701}. In addition, the dimer order and plaquette order can be revealed from the critical behavior of the spin correlations on the different bonds at the quantum phase transition, which are discussed in the Appendix \ref{appa} with more details.

With the coordination number $z=3$ of the 1/5-depleted square lattice, an AKLT state can be form in the spin-3/2 case. However, due to the nature of two-dimensional bipartite lattice, the ground state of pure Heisenberg model without the biquadratic and bicubic terms is more likely to be a N\'eel phase for arbitrary S. Next, we numerically calculate the spin correlations and get the extrapolated magnetic order in the intermediate N\'eel phase. And we also have numerically confirmed that other magnetic orders, such as block AFM, are not the ground state whatever the spin magnitude is. To detect the N\'eel order, we define the squared staggered magnetization as
\begin{equation}
\begin{aligned}
m_s^2=\frac{1}{N^2} \left\langle\left(\sum_{i=1}^N (-1)^i S^z_i\right)^2\right\rangle,
\end{aligned}
\label{ms2}
\end{equation}
where $(-1)^i=\pm1$ is the staggered phase factor according to the N\'eel-type spin configuration as illustrated in Fig. \ref{phase}{\color{blue}(a)}. The finite-size scaling of the staggered magnetization is expected to be of order $O(N^{-1/2})$ \cite{RevModPhys.63.1,PhysRevLett.98.227202}.

\begin{figure}[tb]
	\centering
	\includegraphics[width=0.45\textwidth]{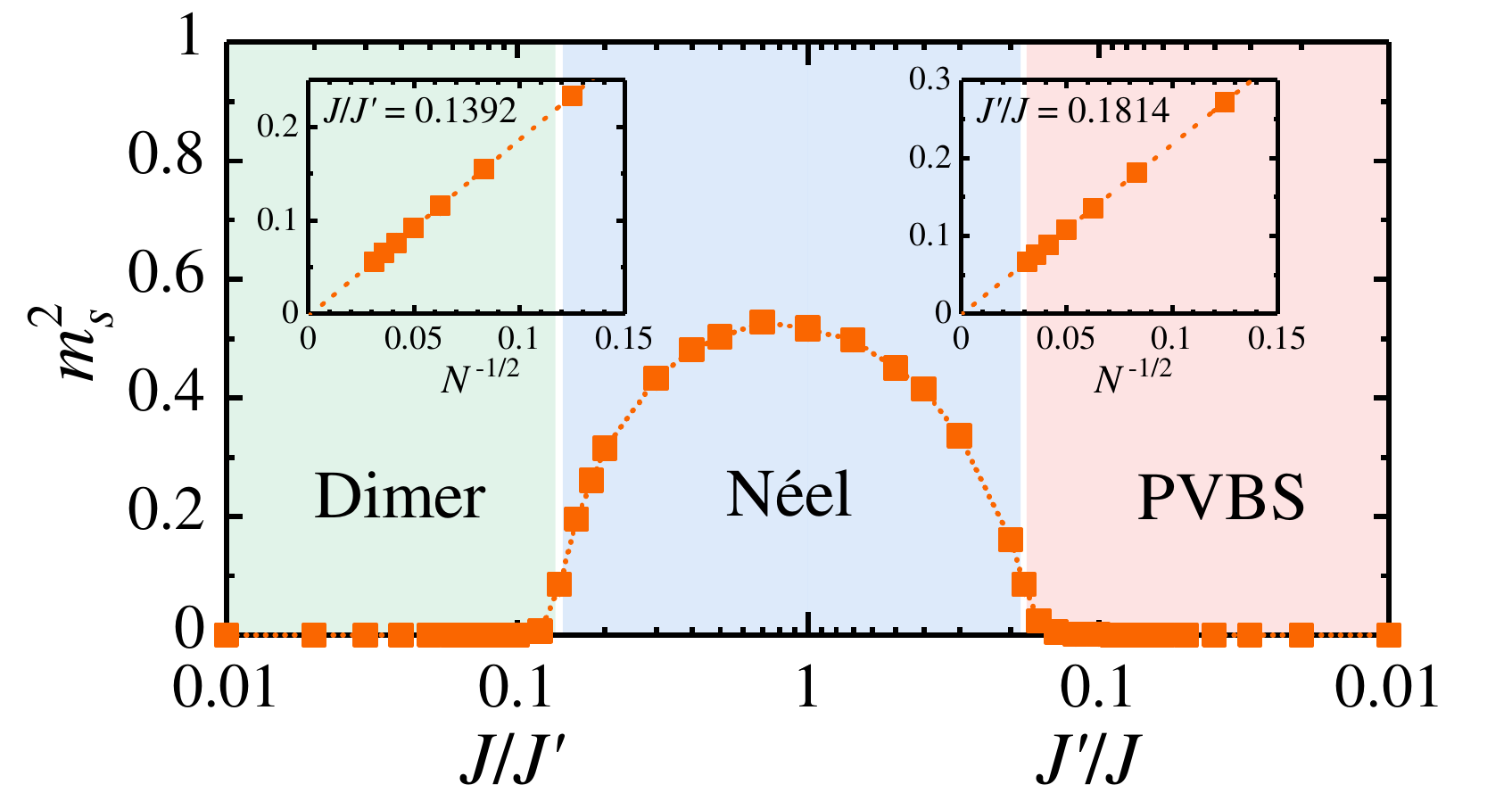}
	\caption{The phase diagram of the $S=3/2$ antiferromagnetic Heisenberg model on the 1/5-depleted square lattice consists of the dimer phase, N\'eel phase and PVBS phase. The squared staggered magnetization $m_s^2$ is obtained from QMC simulations on this lattice with $L=8$ in the whole phase regions. The insets show the finite-size dependence of $m_s^2$ with linear sizes $L=4, 6, 8, 10, 12, 14, 16$ near the quantum critical points $J/J'=0.1392$ (left) and $J'/J=0.1814$ (right), respectively. The error bars are smaller than the symbols.}
	\label{s15neel}
\end{figure}

In Fig. \ref{s15neel}, the squared staggered magnetization $m_s^2$ in the thermodynamic limit is shown versus the coupling ratios $J/J'$ for $J\le J'$ and $J'/J$ for $J\ge J'$, respectively. The squared staggered magnetization gradually decreases to zero as expected in the disordered dimer phase and the PVBS phase. The insets of Fig. \ref{s15neel} further show the finite-size dependence of $m_s^2$ as a function of $N^{-1/2}$ near the quantum critical points, in which total sites like $N=64, 144, 256, 400, 576, 784, 1024$ (where $N=4L^2$) are used to do the extrapolations. The squared staggered magnetization $m_s^2$ is extrapolated to zero in the thermodynamic limit at these two quantum critical points. However, the N\'eel order parameter $m_s^2$ in the intermediate phase as shown in Fig. \ref{s15neel} is enhanced compared to the spin-1/2 case.

In conclusion, we have got the phase diagram of the $S=3/2$ antiferromagnetic Heisenberg model on the 1/5-depleted square lattice, including the dimer phase, the N\'eel phase and the PVBS phase with the quantum critical points $J_c=0.1392(2)$ and $J'_c=0.1814(4)$. These two quantum phase transition belong to the three-dimensional O(3) universality class. We also show the magnetization of the N\'eel phase in the thermodynamic limit.

\subsection{The spin-1 case}
In this section, we explore the ground-state properties of the $S=1$ antiferromagnetic Heisenberg model on the 1/5-depleted square lattice in order to seek rule without further calculations on other higher-spin case. Similar to the spin-3/2 case, we can get the phase boundaries with high accuracy by using large-scale finite-size scaling.

\begin{figure}[tb]
	\centering
	\includegraphics[width=0.45\textwidth]{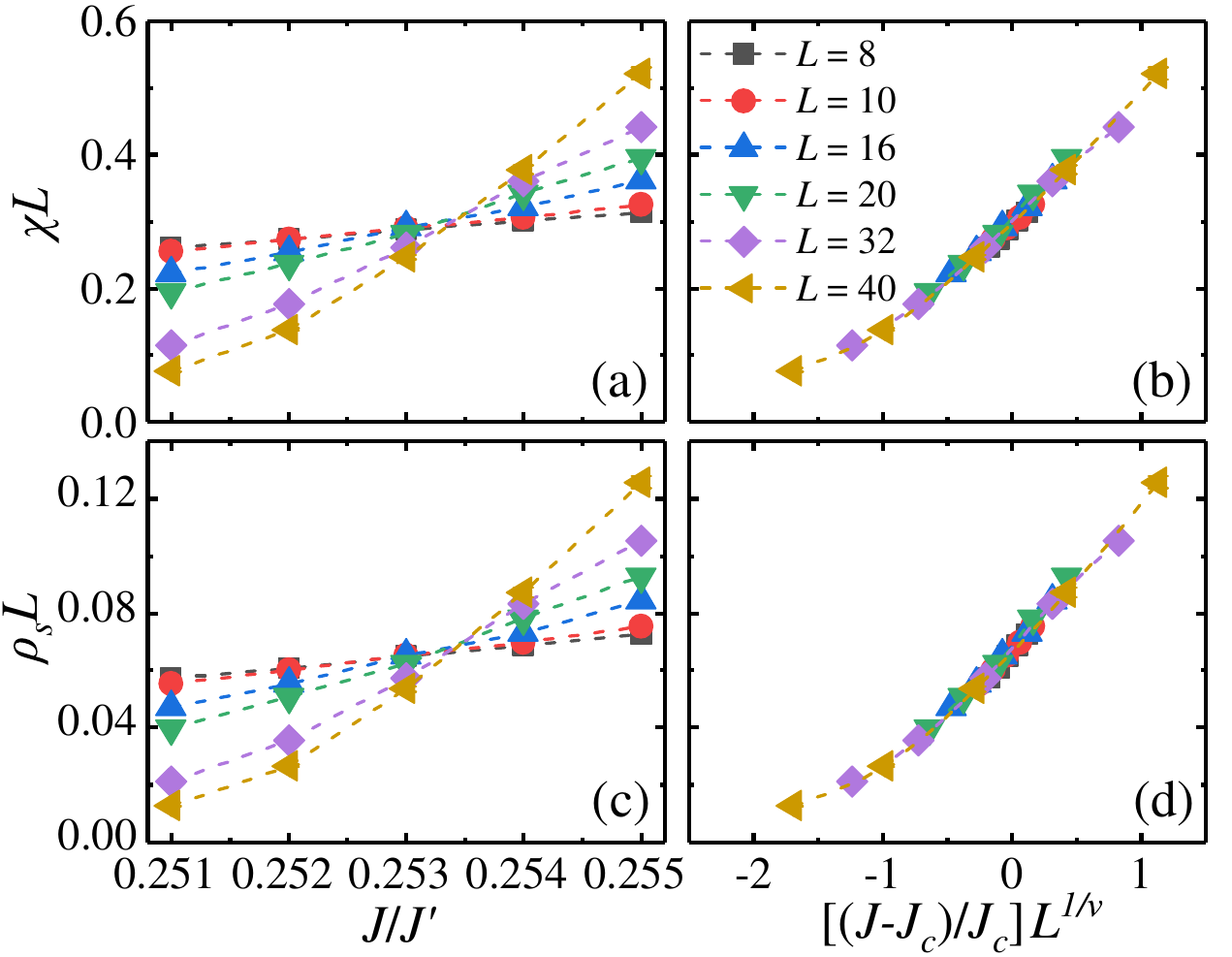}
	\caption{(a),(b) The uniform magnetic susceptibility multiplied by $L$ and (c),(d) the spin stiffness multiplied by $L$ of the $S=1$ antiferromagnetic Heisenberg model on the 1/5-depleted square lattice near the quantum critical point $J_c$. Data collapses are achieved for a fixed standard O(3) value $1/\nu =1.406$ and $J_c=0.2534(4)$.}
	\label{s1qcp1}
\end{figure}

To get the quantum critical points, we also extract the two dimensionless quantities, the $L$-normalized uniform magnetic susceptibility $\chi L$ and the spin stiffness $\rho_s L$ defined in Eqs. (\ref{chi}) and (\ref{rhos}). Figures \ref{s1qcp1}{\color{blue}(a)} and \ref{s1qcp1}{\color{blue}(c)} show the results of $\chi L$ and $\rho_s L$ with various linear sizes $L=8, 10, 16, 20, 32, 40$ near the quantum critical point $J_c$ between the dimer phase and the N\'eel phase. Figures \ref{s1qcp1}{\color{blue}(b)} and \ref{s1qcp1}{\color{blue}(d)} show a good data collapse with a fixed standard O(3) value $1/\nu =1.406$ and an accurate estimate $J_c=0.2534(4)$ according to finite-size scaling hypothesis at criticality. Therefore, the quantum phase transition from the dimer phase to the N\'eel phase is a continuous transition with the critical point $J_c=0.2534(4)$.

\begin{figure}[tb]
	\centering
	\includegraphics[width=0.45\textwidth]{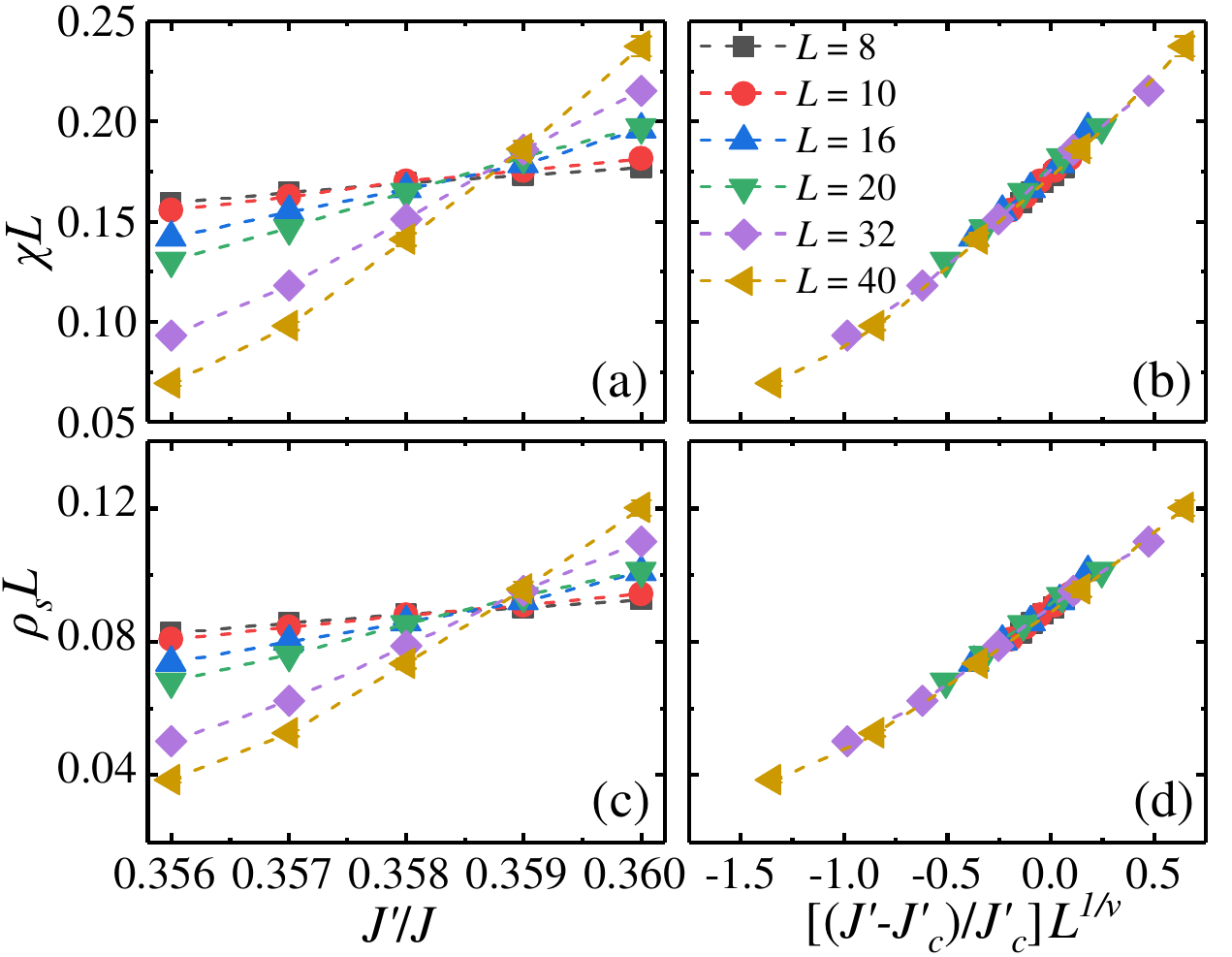}
	\caption{(a),(b) The uniform magnetic susceptibility multiplied by $L$ and (c),(d) the spin stiffness multiplied by $L$ of the $S=1$ antiferromagnetic Heisenberg model on the 1/5-depleted square lattice near the quantum critical point $J'_c$. Data collapses are achieved for a known standard O(3) value $1/\nu =1.406$ and $J'_c=0.3587(4)$.}
	\label{s1qcp2}
\end{figure}

As seen in Figs. \ref{s1qcp2}{\color{blue}(a)} and \ref{s1qcp2}{\color{blue}(c)}, the uniform magnetic susceptibility $\chi L$ and the spin stiffness $\rho_s L$ with different linear sizes $L$ both also cross each other at the quantum critical point $J'_c$ between the N\'eel phase and the PVBS phase in the $S=1$ case. Similarly, we fix the correlation length exponent at the standard O(3) value $1/\nu =1.406$ and perform finite-size data collapse fits to find a precise estimate of $J'_c$ as shown in Figs. \ref{s1qcp2}{\color{blue}(b)} and \ref{s1qcp2}{\color{blue}(d)}. The quantum critical point between the N\'eel phase and the PVBS phase is estimated to be $J'_c=0.3587(4)$. We summarize the accurate estimations of the quantum critical points for $S\ge1/2$, which are listed in Table \ref{table1}.

\begin{table}[b]
	\renewcommand{\arraystretch}{1.5}
	\caption{\label{table1}
		The estimated results of the quantum critical points on the 1/5-depleted square lattice for different spins. The phase boundaries for the spin-1/2 case are quoted from Ref. \cite{PhysRevLett.118.087201}.
	}
	\begin{ruledtabular}
		\begin{tabular}{cll}
			\textrm{Spin $S$}&
			\textrm{$J_c~(J'=1)$}&
			\textrm{$J'_c~(J=1)$}\\
			\colrule	
			1/2 & 0.603520(10) & 0.939512(12) \\
			1 & 0.2534(4) & 0.3587(4) \\
			3/2 & 0.1392(2) & 0.1814(4) \\
		\end{tabular}
	\end{ruledtabular}
\end{table}

\begin{figure}[htb]
	\centering
	\includegraphics[width=0.4\textwidth]{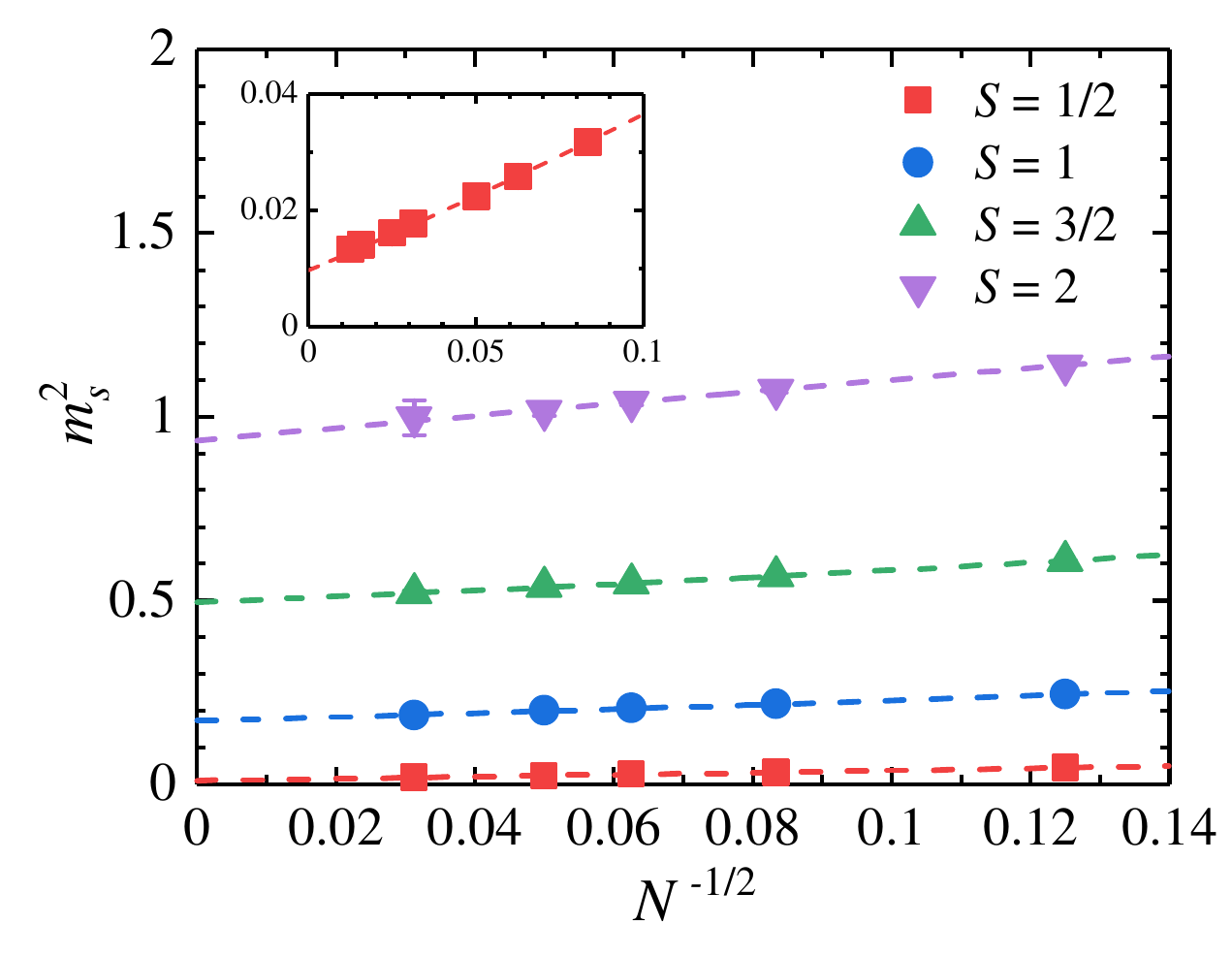}
	\caption{The squared staggered magnetization $m_s^2$ of the antiferromagnetic Heisenberg model on the 1/5-depleted square lattice as a function of $N^{-1/2}$ (where $N=4L^2$) at $J/J'=1$ for different spins $S=1/2, 1, 3/2, 2$. The dashed curves are the second-order polynomial fits. The inset shows more details about the finite-size extrapolation for the spin-1/2 case.}
	\label{allneel}
\end{figure}

Additionally, in order to verify the suppression of quantum fluctuation when $S$ gets larger and goes to the classical Heisenberg limit, we calculate the squared staggered magnetization $m_s^2$ defined in Eq. (\ref{ms2}) with different $S$. Figure \ref{allneel} shows the finite-size extrapolations of $m_s^2$ versus $N^{-1/2}$ on the 1/5-depleted square lattice with different system sizes $N=64, 144, 256, 400, 1024$ at $J/J'=1$. As we expect, the squared staggered magnetization for different spins can be extrapolated to non-zero values in the thermodynamic limit by using the second-order polynomial fits. The finite-size scaling results of the squared staggered magnetization are $m_s^2=0.0097(3), 0.172(6), 0.495(3), 0.935(16)$ for the spin magnitudes $S=1/2, 1, 3/2, 2$, respectively. For the spin-$1/2$ case, the long-range antiferromagnetic order is relatively small due to the quantum fluctuation and depleted characteristics, whereas the antiferromagnetic order shows a considerable increase towards the classical limit as $S$ gets larger.

To conclude, we get the accurate critical points in the phase diagrams of the Heisenberg model on the 1/5-depleted square lattice for various spin magnitudes, which are summarized in Table \ref{table1}. The whole phase diagrams versus the coupling ratios for different spins $S=1/2, 1, 3/2$ are shown in Fig. \ref{phase}{\color{blue}(c)}. A higher proportion of the intermediate N\'eel phase can be found on the lattice with higher spins, which means that the disordered dimer and PVBS phases are originated from quantum nature of magnetic systems.

\subsection{Magnetic excitations}

\begin{figure}[tb]
	\centering
	\includegraphics[width=0.4\textwidth]{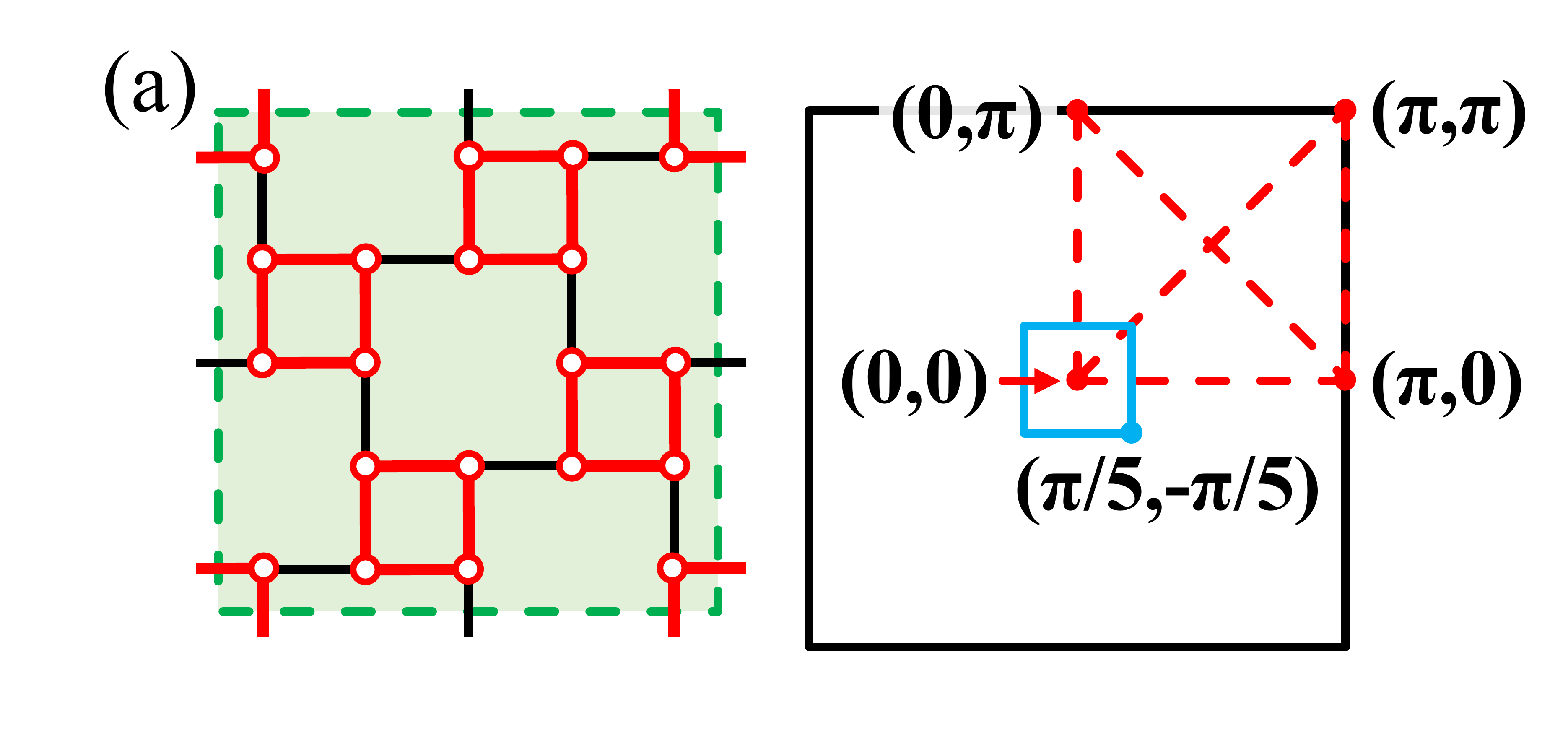}
	\includegraphics[width=0.45\textwidth]{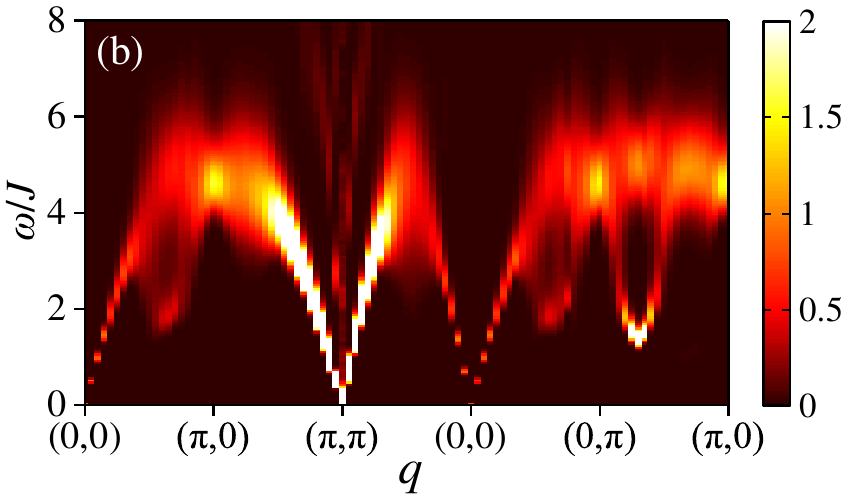}
	\includegraphics[width=0.45\textwidth]{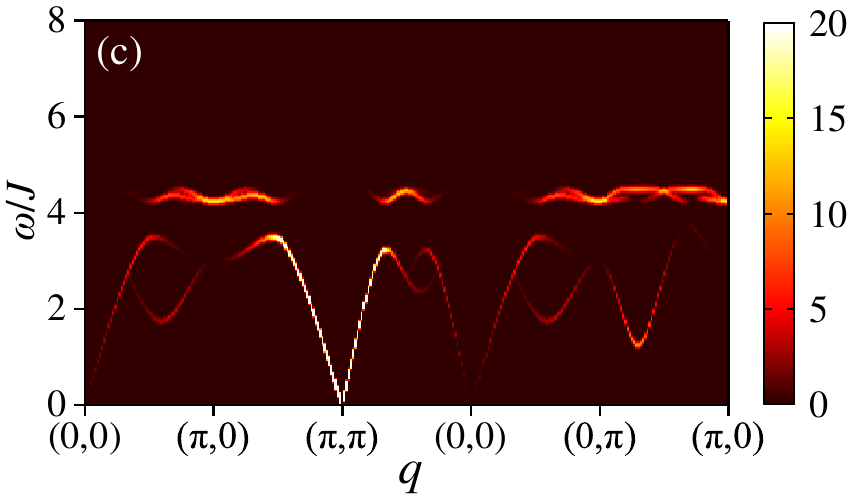}
	\caption{(a) Left panel: The twenty-site supercell (green dashed square) on the 1/5-depleted square lattice. Right panel: The folded Brillouin zone (blue small solid square) and the wave vector path (red dashed lines) in the extended Brillouin zone. (b) and (c) show the dynamic spin structure factor $S^{zz}(\bm{q},\omega)$ of the $S=3/2$ antiferromagnetic Heisenberg model on the 1/5-depleted square lattice at $J/J'=1$ which is in the N\'eel phase. Panel (b) is obtained from QMC simulations and stochastic analytic continuation with linear system size $M=8$ of supercell and $\beta=40$. Panel (c) is obtained from linear spin wave theory.}
	\label{sqwspinw}
\end{figure}

Next, we study the spin excitations of the spin-3/2 Heisenberg antiferromagnet on the 1/5-depleted square lattice. The longitudinal dynamic spin structure factor $S^{zz}(\bm{q},\omega)$ is obtained from the QMC simulations combined with stochastic analytic continuation along the path $(0,0) \rightarrow (\pi,0) \rightarrow (\pi,\pi) \rightarrow (0,0) \rightarrow (0,\pi) \rightarrow (\pi,0) $ of extended Brillouin zone. To make more momentum points fall into that path in the finite-size lattice, we choose the lattices to be the multiples of the supercell shown in Fig. \ref{sqwspinw}{\color{blue}(a)} \cite{PhysRevB.99.085112}.

In Fig. \ref{sqwspinw}{\color{blue}(b)}, we show the dynamic spin structure factor $S^{zz}(\bm{q},\omega)$ of the $S=3/2$ antiferromagnetic Heisenberg model on the 1/5-depleted square lattice with linear system size $M=8$ of supercell [illustrated in the left panel of Fig. \ref{sqwspinw}{\color{blue}(a)}] and $\beta=40$ at the coupling ratio $J/J'=1$, which belongs to the N\'eel phase. And the total lattice size is equal to $N=20M^2$. From the excitation spectra, the gapless Goldstone mode can be found at the wave vector $\bm{q}=(\pi,\pi)$ as we expect. And the spectral weight of the gapless magnon mode nearly vanishes at $\bm{q}=(0,0)$ due to the conservation of $S^z$ \cite{PhysRevB.98.174421,PhysRevLett.122.175701}. Moreover, a novel magnon dispersion occurs between the wave vectors $\bm{q}=(0,\pi)$ and $(\pi,0)$ because of the presence of a magnon pole around $\bm{q}\approx(\pi/5,3\pi/5)$ owing to the depleted characteristic of the lattice and the Brillouin zone folding.

\begin{figure*}[tb]
	\centering
	\includegraphics[width=0.95\textwidth]{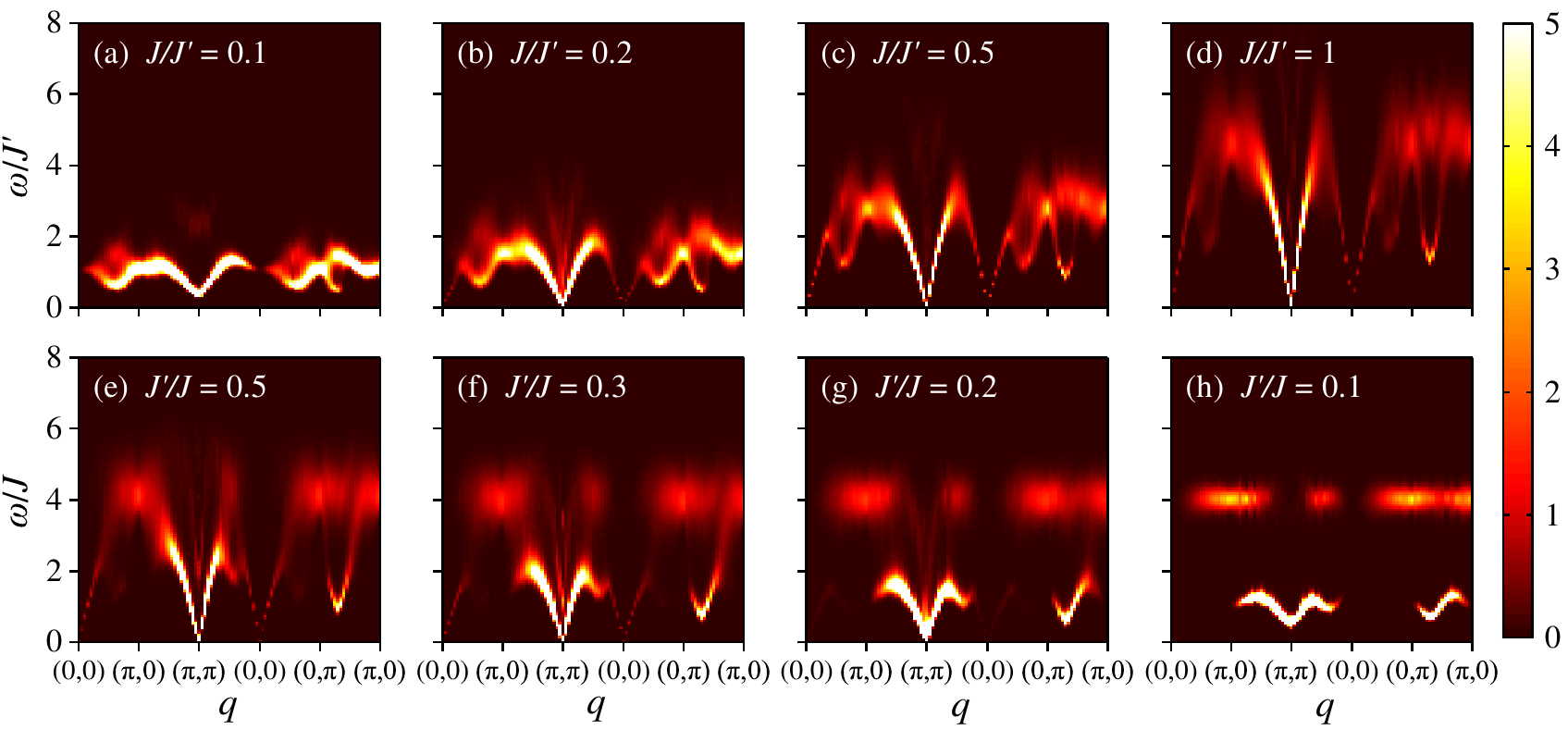}
	\caption{The dynamic spin structure factor $S^{zz}(\bm{q},\omega)$ obtained from QMC simulations and stochastic analytic continuation for the $S=3/2$ antiferromagnetic Heisenberg model on the 1/5-depleted square lattice with linear system size $M=8$ of supercell and $\beta=40$. Here (a) is in the dimer phase, (b)-(g) are in the N\'eel phase, and (h) is in the PVBS phase.}
	\label{sqws15}
\end{figure*}

\begin{figure*}[t]
	\centering
	\includegraphics[width=0.95\textwidth]{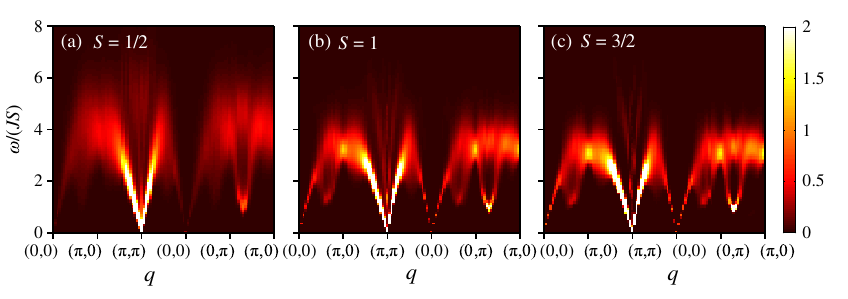}
	\caption{The dynamic spin structure factor $S^{zz}(\bm{q},\omega)$ of the antiferromagnetic Heisenberg model on the 1/5-depleted square lattice with linear system size $M=8$ of supercell and $\beta=40$ at $J/J'=1$ for different spins $S=1/2, 1, 3/2$. The results are obtained from QMC simulations and stochastic analytic continuation.}
	\label{sqwneel}
\end{figure*}

In order to further understand the spin excitations, we show the results of the dynamic spin structure factor $S^{zz}(\bm{q},\omega)$ obtained by linear spin-wave theory as can be seen in Fig. \ref{sqwspinw}{\color{blue}(c)}. A Holstein-Primakoff transformation is performed to bosonize the Hamiltonian, in which the spin operators can be replaced by the boson creation and annihilation operators \cite{PhysRevB.103.064417}:
\begin{equation}
\begin{aligned}
&S^{z}_{i}=S-a^\dagger_{i}a_{i},
&S^{+}_{i}\approx\sqrt{2S}a_{i}, \quad
&S^{-}_{i}\approx\sqrt{2S}a^\dagger_{i}, \\
&S^{z}_{j}=b^\dagger_{j}b_{j}-S,
&S^{+}_{j}\approx\sqrt{2S}b^\dagger_{j}, \quad
&S^{-}_{j}\approx\sqrt{2S}b_{j},
\end{aligned}
\label{boson}
\end{equation}
where $a^\dagger_{i}$, $a_{i}$ ($b^\dagger_{j}$, $b_{j}$) are for up (down) spins as illustrated in Fig. \ref{phase}{\color{blue}(a)}. Then the spin wave dispersions and dynamic spin structure factor $S^{zz}(\bm{q},\omega)$ can be calculated after diagonalizing the Hamiltonian. As shown in Fig. \ref{sqwspinw}{\color{blue}(c)}, the spin wave can nicely capture the low-energy excitation spectra obtained from the QMC simulations. However, there are nearly flat bands in the high energy separating from the low-energy branches in the spin wave dispersions, which is different to the high-energy broad spectrum of the QMC results (see Fig. \ref{sqwspinw}). Here, we provide some potential reasons to account for the high-energy part. Firstly, the stochastic analytic continuation numerical methods have not yet distinguished the multimagnon continua and the single-magnon quite well in a small range of frequencies \cite{arXiv2202.09870}. Secondly, the high-energy continua may be contributed by magnons with strong interactions and even nearly deconfined spinons, which cannot be simply captured by the linear spin wave theory \cite{Nat.Phys.9.435}. And the presence or absence of the nearly-deconfined excitations around $\bm{q}=(\pi,0)$ and $(0,\pi)$ need to be further confirmed.

\begin{figure}[htb]
	\centering
	\includegraphics[width=0.45\textwidth]{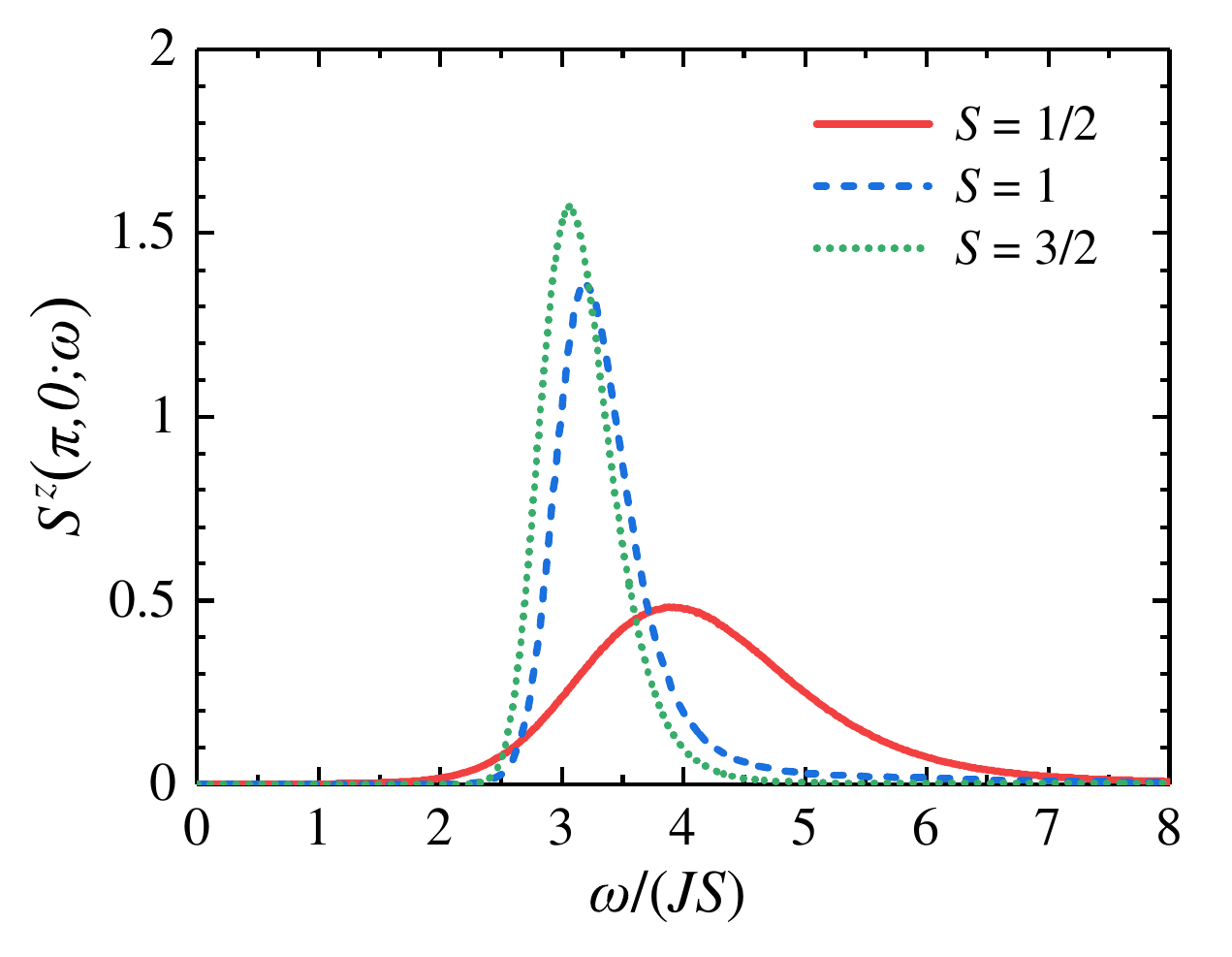}
	\caption{The dynamic spin structure factor $S^{zz}(\bm{q},\omega)$ of the uniform antiferromagnetic Heisenberg model on the 1/5-depleted square lattice with $M=8$ and $\beta=40$ at the wave vector $\bm{q}=(\pi,0)$ for different spins $S=1/2, 1, 3/2$.}
	\label{sqwpi0}
\end{figure}

When the coupling ratio $J/J'$ keep decreasing towards the quantum critical point $J_c=0.1392(2)$, the excitation spectra are gradually pushed to lower energy and trend to be gapped at the wave vectors $\bm{q}=(0,0)$ and $(\pi,\pi)$ as shown in Fig. \ref{sqws15}{\color{blue}(b)-(d)}. In the dimer phase, we choose a representative point $J/J'=0.1$ ($J'=1$) to study the dynamical properties. In Fig. \ref{sqws15}{\color{blue}(a)}, we show the dynamic spin structure factor $S^{zz}(\bm{q},\omega)$ at $J/J'=0.1$ using the QMC simulations and stochastic analytic continuation. The excitation spectrum are gapped as expected. The spin singlets can form on the dimer bonds. And a dimer singlet can be excited to a triplet that can move in the whole lattice, which is a well-known triplon excitation. Further decreasing $J/J'$ to the isolated dimer limit, a nearly flat band can be observed in the spectrum, which is consistent with the exact diagonalization results shown in Appendix \ref{appb}. In Fig. \ref{sqws15}{\color{blue}(e)-(g)}, we also show the evolution of the excitation spectra when $J'/J$ decreases (i.e., $J/J'$ increases) towards the quantum critical point $J'_c=0.1814(4)$. An observable separation process can be found between the low-energy magnon mode and the high-energy continuum. It is worth mentioning that the separation does not occur exactly at the critical point. More discussions in other spin-$S$ system can be found in Appendix \ref{appc}. Similarly, we choose the coupling ratio $J'/J=0.1$ ($J=1$) for the PVBS phase. In Fig. \ref{sqws15}{\color{blue}(b)}, a fully gapped spectrum can also be found, and a prominent triplon mode appears around $\omega=1.0$. In the high-energy part, there is another excitations separating from the low-energy triplon mode in the PVBS phase \cite{PhysRevB.99.085112}, which can be captured by the energy spectrum of the isolated plaquette.

In addition, we also extract the excitation spectra of the spin-1/2 and spin-1 cases aiming to compare the similarities and differences between the spin-3/2 and lower-spin case. As shown in Fig. \ref{sqwneel}, the dynamic spin structure factor $S^{zz}(\bm{q},\omega)$ at $J/J'=1$ are shown for different spins $S=1/2, 1, 3/2$, which all belong to the N\'eel phase [see Fig. \ref{phase}{\color{blue}(c)}]. The overall shapes of the spin-1/2 and spin-1 excitation spectra are similar to the spin-3/2 case due to the existence of the gapless Goldstone mode in the N\'eel phase. The more detailed results of the dynamic spin structure factor $S^{zz}(\bm{q},\omega)$ at the wave vector $\bm{q}=(\pi,0)$ are shown in Fig. \ref{sqwpi0}. A broader high-energy continuum can be found in the excitation spectra of the lower-spin case, especially for $S=1/2$, which may indicate the presence of the nearly deconfined spinons \cite{PhysRevB.88.144414}. However, for the higher-spin case, the broad continuum disappears, which may be due to the confinement of spinons in the classical limit $S\rightarrow \infty$.

\section{CONCLUSION}
\label{conclu}
In this work, we have revealed the phase diagram of the spin $S>1/2$ antiferromagnetic Heisenberg model on the 1/5-depleted square lattice. By using the extensive finite-size scaling of the quantum Monte Carlo results, we obtain the accurate quantum critical points, and numerically verify that the continuous quantum phase transitions belong to the three-dimensional O(3) universality class.

To generalize to other higher-spin case, we have representatively study the ground-state properties of 1/5-depleted square-lattice Heisenberg model for the spin-1 and spin-3/2 case. According to the QMC results, when the spin magnitude increases, the magnetic order enhances, and the region of N\'eel phase extends to a larger area. Thus, in the higher spin case, very weak interactions between the plaquettes (or dimers) can give rise to the N\'eel phase. In other words, quantum fluctuation becomes very weak to suppress the long-range antiferromagnetic order as the spin magnitude increases.

Moreover, we have studied the dynamical properties of the $S=3/2$ Heisenberg model versus the coupling ratio on the 1/5-depleted square lattice. The dynamic spin structure factor $S^{zz}(\bm{q},\omega)$ is well extracted by using stochastic analytic continuation of the imaginary-time correlation function obtained from the QMC simulations. In the dimer phase and the PVBS phase, the low-energy excitations are numerically verified as the gapped triplons. In the N\'eel phase, there are well-defined gapless Goldstone modes (magnons) at the wave vectors $\bm{q}=(\pi,\pi)$ and $(0,0)$, which is consistent with the linear spin wave theory. And the depleted characteristic of the lattice and the Brillouin zone folding give rise to a magnon pole around $\bm{q}\approx(\pi/5,3\pi/5)$. What's more, the evolution of the separation between the low-energy and high-energy spectra can be well studied owing to the two triplet excitations in the isolated PVBS limit. Finally, we have also calculated the dynamic spin structure factor $S^{zz}(\bm{q},\omega)$ for the spin-1/2 and spin-1 cases. The excitation spectra of the lower-spin case show a broader continuum, especially at $(\pi,0)$ and $(0,\pi)$, suggesting that the nearly deconfined spinons may exist.

The spin-$S$ Heisenberg model on the 1/5-depleted square lattice can be simulated with ultracold atoms in optical lattices in the future or be synthesized in more Mott insulators with multiorbitals. Our numerical results can provide guidance for realizing different phases in this model. And the excitation spectra with different spin magnitudes provide a playground for studying the gapped triplons, the magnon and possible nearly deconfined spinon, which can be identified in the inelastic neutron scattering experiments. In addition, adding the biquadratic and bicubic interactions in the $S=3/2$ case can induce an AKLT phase and some other phases, which is still worthy to be studed by the density matrix renormalization goup and tensor network in future.

\begin{acknowledgments}
The authors would like to thank Hui Shao and Shangjian Jin for helpful discussions. This work is supported by NKRDPC-2017YFA0206203, NKRDPC-2018YFA0306001, NSFC-11804401, NSFC-11974432, NSFC-11832019, GBABRF-2019A1515011337, Leading Talent Program of Guangdong Special Projects (201626003), Shenzhen Institute for Quantum Science and Engineering (Grant No. SIQSE202102), and Fundamental Research Funds for the Central Universities, Sun Yat-sen University (Grant No.2021qntd27).
\end{acknowledgments}

\appendix
\section{SPIN CORRELATIONS ON THE DIMER AND PLAQUETTE BONDS}
\label{appa}

\begin{figure}[htb]
	\centering
	\includegraphics[width=0.45\textwidth]{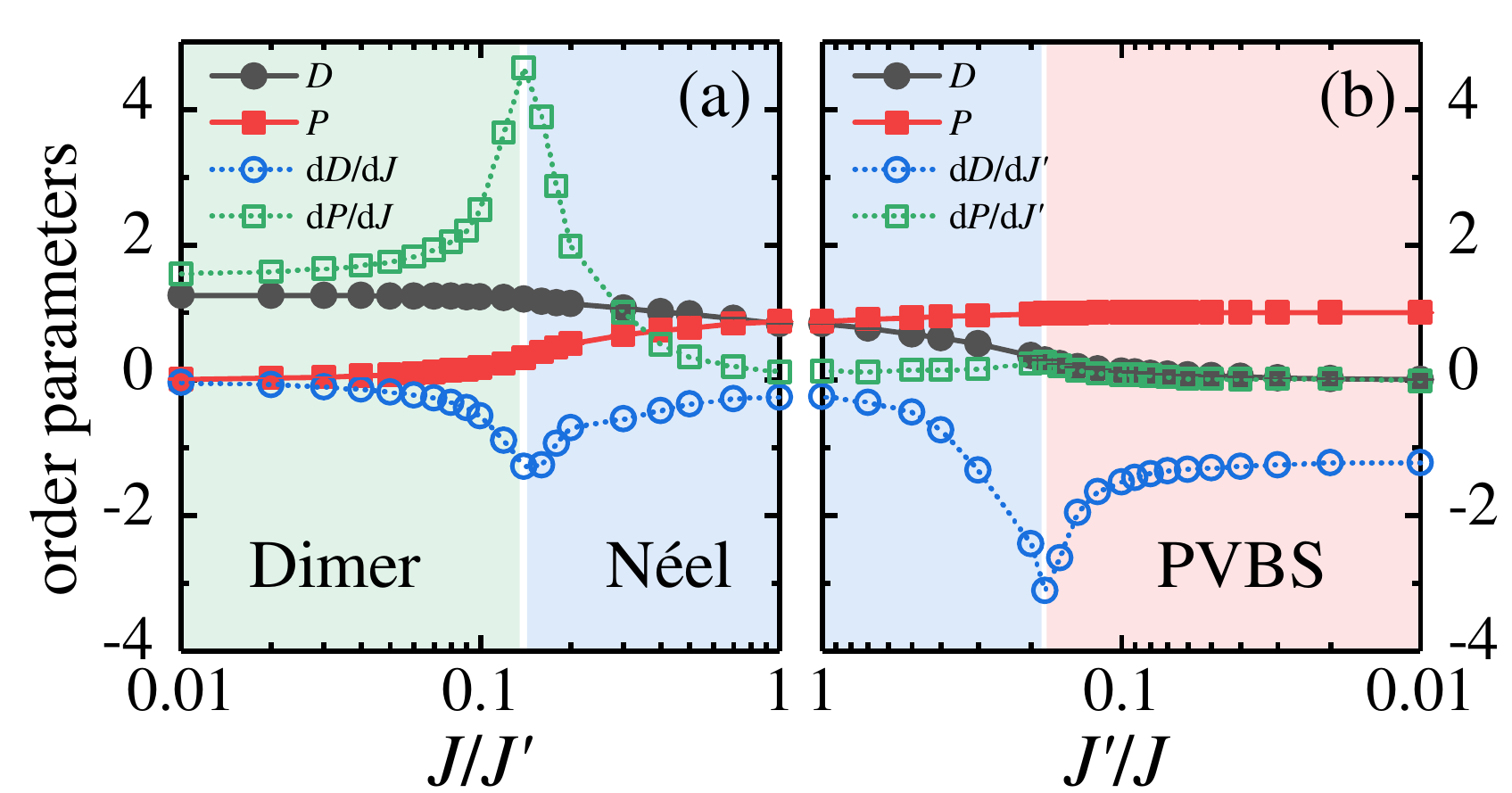}
	\caption{The dimer order parameter $D$ and the plaquette order parameter $P$ are calculated by QMC simulations with $L=8$. The phase boundaries deduced from the first derivatives of $D$ and $P$ with respect to the coupling ratios agree with our finite-size scaling results in the main text.}
	\label{s15phase}
\end{figure}


In this section, we discuss the dimer order and the plaquette order of the spin-3/2 antiferromagnetic Heisenberg model on the 1/5-depleted square lattice. In order to detect the dimer and the plaquette orders in different phases, we can use the spin correlations of inter-plaquette and intra-plaquette nearest-neighbor bonds as order parameters. The dimer order parameter can be defined as
\begin{equation}
\begin{aligned}
D=-\frac{2}{N}\sum_{\langle ij \rangle'}\langle S_i^zS_j^z\rangle,
\end{aligned}
\label{dimer}
\end{equation}
and the plaquette order parameter is defined as
\begin{equation}
\begin{aligned}
P=-\frac{1}{N}\sum_{\langle ij \rangle}\langle S_i^zS_j^z\rangle
\end{aligned}
\label{plaque}
\end{equation}
Here, $\langle ij \rangle'$ denotes the dimer bonds, and $\langle ij \rangle$ denotes the intra-plaquette bonds \cite{PhysRevB.85.134416}. The parameter $N/2$ represents the number of the dimer bonds. We introduce minus signs in Eqs. (\ref{dimer}) and (\ref{plaque}) on account of the antiferromagnetic couplings. These two order parameters are also the hallmarks of first derivations of ground-state energy with respect to $J$ and $J'$ according to Hellmann–Feynman theorem.

Figure \ref{s15phase} shows the dimer order parameter $D$ and the plaquette order parameter $P$ versus the coupling ratios. As is expected, the dimer order parameter $D$ decreases and the plaquette order parameter $P$ increases gradually as the coupling ratio $J/J'$ is increased. Moreover, the first derivative of the plaquette order parameter $P$ with respect to $J/J'$ reaches a local maximum at the quantum critical point between the dimer phase and the N\'eel phase, which means a rapidly decreasing $P$ when entering the dimer phase. Similarly, as shown in Fig. \ref{s15phase}{\color{blue}(b)}, the plaquette order parameter $P$ dominates, and the dimer order parameter $D$ is reduced to near zero quickly in the PVBS phase.

\section{ISOLATED DIMER AND PLAQUETTE}
\label{appb}

\begin{figure}[htb!]
	\centering
	\includegraphics[width=0.45\textwidth]{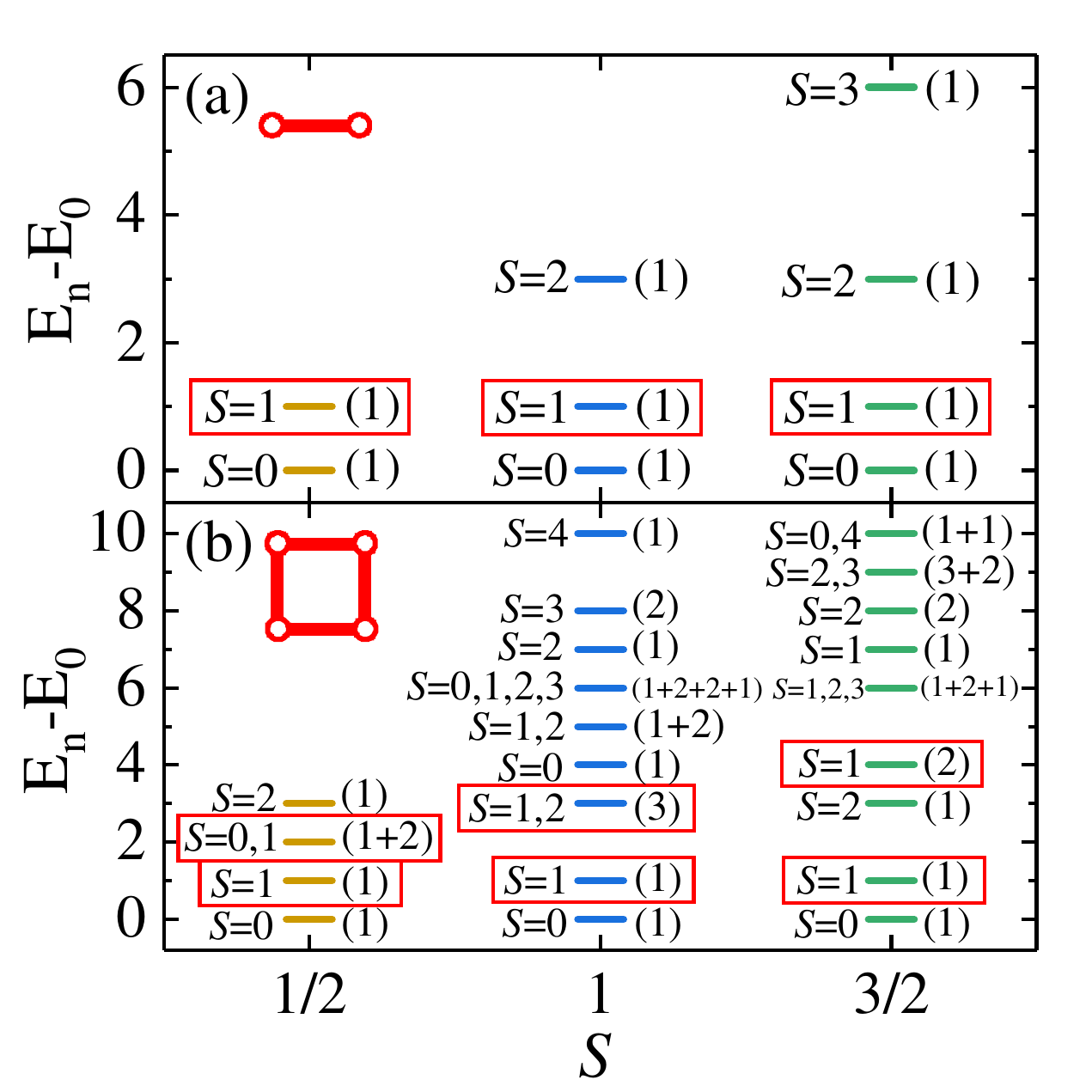}
	\caption{Excitation Spectra of (a) two-site and (b) four-site spin-$S$ Heisenberg model in the $M_z=0$ sector. We show the magnitude of the total spin angular momentum $\sqrt{S(S+1)}\hbar$ in the left of energy levels and the degeneracy in the right of energy levels. The red rectangle boxes represent the triplet excitations, which have nonzero weight in the $S^{zz}(\bm{q},\omega)=\pi\sum_{n}|\langle n|S^z_{\bm{q}} |0\rangle|^2\delta[\omega-(E_n-E_0)]$.}
	\label{edresult}
\end{figure}

The model we study has two limits with isolated dimer and plaquette. Here we show the excitation spectra of two-site and four-site Heisenberg models of spin-1/2,1 and 3/2 in the $M_z=0$ sector using exact diagonalization, where $M_z$ is the eigenvalue of the total spin component along the $z$-th axis (see Fig. \ref{edresult}). In dimer case, there is one triplet excitation above the singlet ground state, and the excitation energy is equal to $J'=1$. In the plaquette case, there are two triplet excitations that contribute to the dynamic spin structure factors in Fig. \ref{sqws15} of the main text.

\begin{figure*}[htb]
	\centering
	\includegraphics[width=0.95\textwidth]{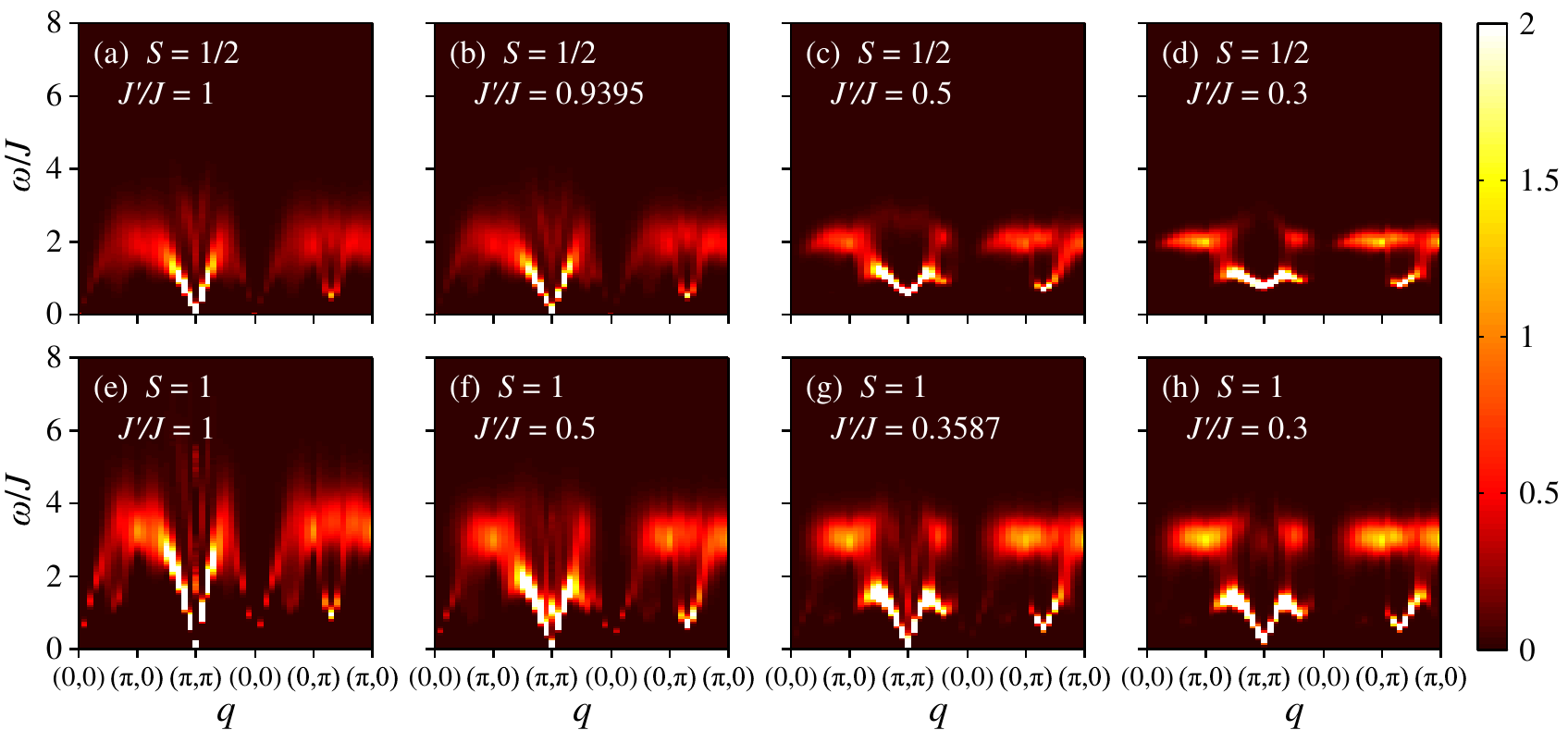}
	\caption{The dynamic spin structure factor $S^{zz}(\bm{q},\omega)$ of the $S=1/2$ and $S=1$ antiferromagnetic Heisenberg model on the 1/5-depleted square lattice with linear system size $M=4$ of supercell and $\beta=20$. Panels (a), (e) and (f) are in the N\'eel phase, (b) and (g) are close to the quantum critical point $J'_c$, and (c), (d) and (h) are in the PVBS phase.}
	\label{cfig}
\end{figure*}

\section{SEPARATION PROCESS BETWEEN THE LOW-ENERGY AND HIGH-ENERGY EXCITATIONS}
\label{appc}

In this section, we further study the dynamical evolutions of the $S=1/2$ and $S=1$ antiferromagnetic Heisenberg model on the 1/5-depleted square lattice versus the coupling ratio $J'/J$. As shown in Fig. \ref{cfig}, the dynamic spin structure factor $S^{zz}(\bm{q},\omega)$ is obtained from QMC calculations and stochastic analytic continuation with linear system size $M=4$ of supercell and $\beta=20$. And we can find that the separation processes of the excitation spectra occur mainly between the coupling ratios $J'/J=0.3$ and $J'/J=0.5$. Similar to the $S=3/2$ case in the main text, the separations between the low-energy and high-energy excitations do not always occur with the quantum phase transition synchronously, particularly in the $S=1/2$ case.

\bibliography{reference}

\end{document}